\documentclass[aps,prb,twocolumn,showpacs,amsmath,amssymb]{revtex4}
\usepackage{epsfig}
\usepackage{dcolumn}
\usepackage{bm}
\usepackage{amsmath}
\usepackage{amsfonts}
\usepackage{amssymb}
\usepackage{graphicx}%
\newcommand{\beq}{\begin{eqnarray}}
\newcommand{\eeq}{\end{eqnarray}}

\begin{document}

\title{Voltage profile and four terminal resistance of an interacting quantum wire.}
\author{Liliana Arrachea}
\affiliation{Departamento de F\'{\i}sica de la Materia Condensada and BIFI, Universidad
de Zaragoza,  Pedro Cerbuna 12, 50009 Zaragoza, Spain.\\
Departamento de F\'{\i}sica, FCEyN, UBA, Pabell\'on 1 Ciudad Universitaria,
1428 Buenos Aires, Argentina.}
\author{Carlos Na\'on and Mariano Salvay}
\affiliation{Departamento de F\'{\i}sica, Facultad de Ciencias
Exactas, Universidad Nacional de La Plata and IFLP-CONICET, CC 67,
 1900 La Plata, Argentina.}
\begin{abstract}
We investigate the behavior of the four-terminal resistance
$R_{4pt}$ in a quantum wire described by a Luttinger liquid
in two relevant situations: (i) in the presence of a single
impurity within the wire and (ii) under the effect of asymmetries introduced by
disordered voltage probes. In the first case, interactions leave a
signature in a power law behavior of $R_{4pt}$ as a function of
the voltage $V$ and the temperature $T$. In the second case
interactions tend to mask the effect of the asymmetries. In both
scenarios the occurrence of negative values of $R_{4pt}$ is
explained in simple terms.
\end{abstract}
 \maketitle

The proposal of a fundamental relation between the two terminal conductance $G$ and
the universal quantum $G_0=e^2/h$ is one of the milestones of electronic quantum
transport in mesoscopic systems \cite{conwinonin1,conwinonin2}. For
 non-interacting electrons,  such a  relation explicitly reads $G=n
G_0$, being $n$ the number of electronic channels \cite{conwinonin2}. The same relation
has been later theoretically and experimentally probed to be also valid in
the case of wires of interacting electrons of
finite length ideally attached to  non-interacting leads
\cite{conwiint,conwiex}.

Experiments in single-wall nanotubes (SWNTs), ropes of SWNTs and
also in multiple-wall nanotubes (MWNTs) \cite{condnanot12} have,
instead, identified that the tunneling conductance to metallic
contacts, $G_t$  follows a power law behavior with the voltage $V$
$G_t \propto V^{\alpha}$ at low temperature $T$, and $G_t \propto
T^{\alpha}$ for low $V$. The exponent ${\alpha}$ being a function
of the forward electron-electron (e-e) interaction $g$. These
features can be understood within the framework of a Luttinger
liquid (LL) theory by means of
 theoretical treatments \cite{conlut} going beyond linear response in $V$ and $T$.

Recently, a combined structure of MWNTs and SWNTs has been used to
analyze the behavior of the four-point resistance $R_{4pt}$ of a
SWNT \cite{nano4pt}. The total resistance of a mesoscopic system
in a two terminal setup contains the
component $1/G_0$
due to the coupling to the
reservoirs. Instead,
$R_{4pt}$  is expected to characterize the genuine resistance of the sample.
For non-interacting electrons at low $V$ and
zero temperature, B\"uttiker has elaborated the concept of the
multi-terminal resistance within scattering-matrix theory (SMT)
\cite{but4pt}, emphasizing the role of the symmetries. Although a
naive expectation would be $R_{4pt} \geq 0$, this theory predicts also the
possibility of $R_{4pt} < 0$ as a consequence of quantum
interference effects. This remarkable feature has been
experimentally observed
 \cite{nano4pt,pict}.

\begin{figure}
\includegraphics[width=0.9\columnwidth,clip]{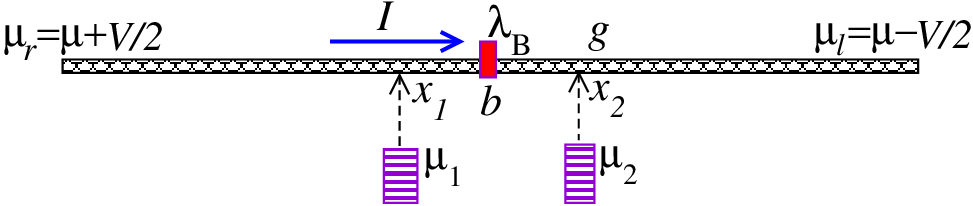}
\caption{\label{fig1}
 Sketch of the setup: A voltage $V$ is imposed on a Luttinger
liquid with a backscattering impurity of strength $\lambda_B$,
through the chemical potentials for the left and right movers:
$\mu_{r,l}=\mu \pm V/2$. Two voltage probes are connected at the
positions $x_1,x_2$. The corresponding chemical potentials
$\mu_{1,2}$ are fixed by the condition of zero current through the
contacts.}
\end{figure}

While the consequences of  elastic scattering due to
impurities can be analyzed in terms of non-interacting electrons,
 the role of the e-e interaction
in the behavior of $R_{4pt}$ remains an open question. The proper
evaluation of this quantity implies dealing with a multi-terminal
setup as the one of Fig. \ref{fig1}, which is difficult to
implement within theoretical approaches like those of Refs.
\onlinecite{conlut,egger,dolc}. Previous multi-terminal treatments
in LL rely in the effective reduction to a non-interacting model
by recourse to a Hartree-Fock decoupling of the interaction term
\cite{lao} or focus in linear response in $V$ \cite{chamon}. In
this work we use Keldysh non-equilibrium Green's functions, which
is a convenient framework to tackle multi-terminal geometries by
exactly treating the e-e interaction while going beyond linear
response. We analyze two relevant ingredients giving rise to a
non-trivial behavior of the local potential $\mu_j$ and
$R_{4pt}$: (i) the presence of an impurity in the wire and (ii) a
clean wire probed by disordered leads with asymmetric densities of
states.

 We consider the setup sketched
in Fig. \ref{fig1}, which consists in a quantum wire described by
a Tomonaga-Luttinger model  under
the influence of a voltage $V$. Two reservoirs are in contact with
the wire at the points $x_1,x_2$ through very weak (``non invasive'')
tunneling constants,
 their chemical potentials $\mu_1, \mu_2$
satisfy the condition of a vanishing current through the ensuing
 contacts. In this way, a continuous current $I$ flows through the wire,
and
\begin{equation} \label{r4pt}
\frac{R_{4pt}}{R_{2pt}}=\frac{\mu_1-\mu_2}{V}.
\end{equation}
 In the case of an ideal and clean wire with identical non invasive probes,
a simple analysis of the symmetries of the setup leads to
 $R_{4pt}=0$.
The full system  is described by the action $
S = S_{wire}+ S_{imp} + S_{res} + S_{cont}$,
where $S_{res}$ describes the two reservoirs that constitute the voltage probes
and $S_{wire}$ is
expressed in terms of right ($r$) and left ($l$) movers (in our  system of units
$\hbar=v_F=e=1$):
\begin{eqnarray}
&S_{wire}&= \int dx\,dt\,\{\psi_r^{\dagger}[i(\partial_t +
\partial_x) - \mu_r]
\psi_r+\nonumber\\
& &+\psi_l^{\dagger}[i(\partial_t -\partial_x) - \mu_l] \psi_l - g
[\psi_r^{\dagger} \psi_r+\psi_l^{\dagger} \psi_l]^2 \} ,
\end{eqnarray}
where $g$ is the Luttinger e-e interaction in the
forward channel while $\mu_r=\mu+V/2$ and $\mu_l=\mu-V/2$.
The effect of the impurity is contained in the back-scattering
interaction:
\begin{eqnarray}
 S_{imp} &=& \lambda_B \int dx\,dt\,\delta(x-b)\,
[e^{-2ik_F x}\psi_{r}^{\dagger} \psi_{l} + H.c].
\end{eqnarray}
The term $S_{cont}$
represents the tunneling between the reservoirs and the
wire:
\begin{eqnarray}
S_{cont} &=& \sum_{j=1,2, \alpha, \beta=l,r}\, \,\int dx\,dt\,
 w_j \delta(x-x_j) \times \nonumber\\ & & [e^{\mp i(k_F+k'_F)
x}\psi_{\alpha}^{\dagger} \chi_{\beta,j}  + H.c],
\end{eqnarray}
where the fields $\chi_{\alpha,j}^{\dagger}$, with $\alpha,
\beta=l,r$, $j=1,2$ denote degrees of freedom of the reservoirs.
The upper and lower sign corresponds to $\alpha=r, l$,
respectively. For simplicity, the dependence of the
fields on $x$ and $t$ has been omitted in the above equations. At
this point we carry out the gauge transformation
$\psi_{l,r}^{\dagger}(x) \rightarrow e^{i \pm k_F x}
\psi_{\alpha}^{\dagger}(x)$ where $k_F$ is the Fermi vector of the
electrons in the wire (a similar transformation involving $k'_F$
is implemented with the fields $\chi_{\alpha,j}^{\dagger}$).

The tunneling currents through the contacts to the probes read
\begin{eqnarray} \label{ij}
I_j = i w_j \sum_{\alpha, \beta=l,r} \langle
\chi_{j,\alpha}^{\dagger}(x_j,t) \psi_{\beta}(x_j,t)- H.c \rangle.
\end{eqnarray}
In what follows, we evaluate the currents $I_j$ up to the first order of perturbation
theory in the tunneling
amplitudes. This procedure
is appropriate
 in the limit of weak $w_j$, which is a reasonable assumption
for measurements of $R_{4pt}$ with `non-invasive' probe leads \cite{nano4pt}.
Within this lowest order of perturbation theory:
\begin{eqnarray} \label{ijp}
I_j & = & 2  |w_j|^2 \sum_{\alpha, \beta =l,r}
\int_{-\infty}^{+\infty} \frac{d \omega}{2 \pi}
[G_{\alpha, \beta}^>(x_j,x_j;\omega)G_{j}^<(\omega)\nonumber \\
& & -G_{\alpha, \beta}^<(x_j,x_j;\omega)G_{j}^>(\omega)],
\end{eqnarray}
where $G_{\alpha,\beta}^{<,>}(\omega)$ are the Fourier transforms
with respect to $t-t'$ of the lesser and bigger Green's functions
$G_{\alpha,\beta}^{<}(x,x';t-t')= i \langle
\psi_{\beta}^{\dagger}(x',t') \psi_{\alpha}(x,t) \rangle$,
$G_{\alpha, \beta }^{>}(x,x';t-t')= - i \langle \psi_{\alpha}(x,t)
\psi_{\beta}^{\dagger}(x',t')  \rangle$, corresponding to the wire
{\em uncoupled from the probes}, while $G_{j}^{>,<}(\omega)  =
\lambda^{>,<}_{j}(\omega) \rho_{j } (\omega)$ are the Green's
functions of the  uncoupled probe reservoirs, with
$\lambda^<_{j}(\omega)= i f(\omega - \mu_{j}) $,
$\lambda^>_{j}(\omega)= -i [1- f(\omega - \mu_{j}) ]$,
 and $\rho_j(\omega)$ the ensuing density of states, which we assume to be identical
for the two kinds of movers within these systems.
The chemical potentials of the probes, $\mu_j$  must be fixed
to satisfy $I_j=0$. Notice that, within this order of perturbation theory, the
effect of the two probes is completely uncorrelated from one another, since interference
terms between the probes and resistive effects involve second order processes in $w_j$
\cite{sec}.

We now turn to analyze the first ingredient of interest, namely,
the effect of an impurity in the wire. In terms of our model, this
corresponds to consider a finite $\lambda_B$. We also consider a
simple model for the probes, with a constant density of states
$\rho_j(\omega)=\rho_0$.
 While the Green's functions of the uncoupled homogeneous
interacting wire are known \cite{medvoit}, the evaluation of the
corresponding functions in the presence of a backscattering center is a non-trivial
task. Below, we indicate the lines we have followed in order to evaluate them
up to the first order of perturbation
theory in $\lambda_B$. The expressions for the
 lesser and bigger
 Green's functions cast:
\begin{eqnarray}
G^{>,<}_{\alpha \beta }(x,x',\omega) & = & \delta_{\alpha \beta}
\lambda^{>,<}_{\alpha}(\omega)
\rho_{0,\alpha } (x-x', \omega)+\nonumber \\
&  & \delta_{\overline{\alpha} \beta} \lambda_B
\{ \lambda^{>,<}_{\alpha}(\omega) \rho_{0,\alpha } (x-b, \omega) \nonumber \\
& & \times
[G^{R}_{0, \beta } (x'-b, \omega)]^*  +
\lambda^{>,<}_{\beta}(\omega) \nonumber \\
& &
\times G^{R}_{0,\alpha } (x-b, \omega)
\rho_{0, \beta } (b-x', \omega) \},
\end{eqnarray}
with $\overline{l}=r, \overline{r}=l$,
$\lambda^<_{\alpha}(\omega)= i f(\omega - \mu_{\alpha}) $,
$\lambda^>_{\alpha}(\omega)= -i [1- f(\omega - \mu_{\alpha}) ]$.
The spectral density
\begin{eqnarray}
\rho_{0,\alpha } (x, \omega+\mu_{\alpha}) &=& C_{\psi}
\exp{[\mp i(\frac{\omega}{v} -k_F)x]} \nonumber \\
& & \times
|\omega|^{2\gamma}
\phi(\gamma,2\gamma+1,\pm 2i x \frac{\omega}{v}),
\end{eqnarray}
corresponds to the clean LL uncoupled from the probes, where
$v=\sqrt{1+2g/\pi}$ is the renormalized Fermi velocity and
$\phi(a,b;c)$ is Kummer's hypergeometric function.  The exponent
$\gamma=(K+K^{-1}-2)/4$, ($K=1/v$) is determined by  $g$. The
retarded Green's functions are defined from the Kramers-Kronig
relation, being $\rho_{0,\alpha } (x, \omega)= -2
\mbox{Im}[G^{R}_{0,\alpha } (x, \omega)]$. In order to perform
numerical computations we introduce an energy cutoff $\Lambda$ by
replacing $\rho_{0,\alpha } (x, \omega+\mu_{\alpha})\rightarrow
\Theta(|\omega|-\Lambda)\rho_{0,\alpha } (x,
\omega+\mu_{\alpha})$. Therefore the constant $C_{\psi}$ is a
function of $\Lambda$ which is determined by the sum rule
$\int_{-\infty}^{+\infty} d \omega \rho_{\alpha,0}(0,\omega) = 2
\pi $.

\begin{figure}
\includegraphics[width=0.9\columnwidth,clip]{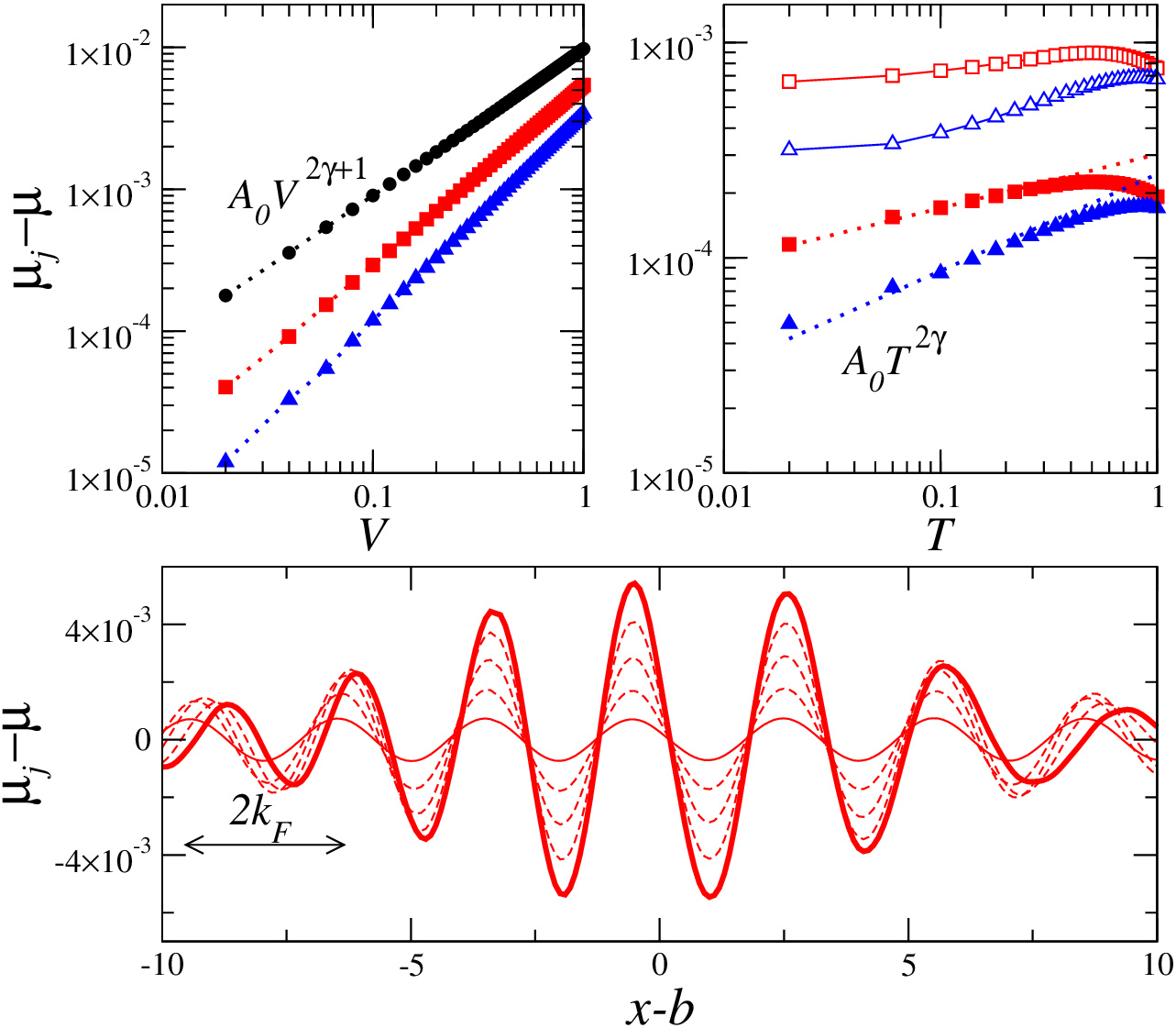}
\caption{\label{fig2} Behavior of the local potential $\mu_j$ for a wire with $\lambda_B=0.1$
as a function of the bias $V$ for $T=0$ and $x_j=-0.2$ (upper left panel) and as a function of
$T$ for $V=0.2$ and $x_j=-0.2$ (right panel).
Circles, squares and triangles correspond to $K=1,0.5,0.4$, respectively. Solid and open symbols correspond to $V=0.05,0.2$, respectively (right panel). The fits with power laws
are shown in dotted lines.
Lower panel:
$\mu_j$ as a function of $x-b$ for  $V=0.2,\ldots, 1$ and $K=0.5$.
}
\end{figure}

Substituting the above expressions in (\ref{ijp}) gives the
following result for the currents through the contacts:
\begin{eqnarray}\label{ij01}
I_j(x_j)& = & 2 |w_j|^2
\sum_{\alpha=l,r} \int_{-\infty}^{\infty} \frac{d\omega}{2\pi}  [f(\omega- \mu_{j})-f(\omega- \mu_{\alpha})] \nonumber \\
& & \times \rho_j(\omega)
\rho^{eff}_{\alpha}(x_j, \omega),
\end{eqnarray}
being
\begin{eqnarray}\label{rhoef}
{\rho}^{eff}_{\alpha}(x_j,\omega) &=& \rho_{0,\alpha } ( 0,\omega)+
2 \lambda_B \mbox{Re}\{
 \rho_{0,\alpha } (x_j-b, \omega)  \nonumber \\
& &
\times [G^{R}_{0, \overline{\alpha} } (x_j-b, \omega)]^* \},
\end{eqnarray}
the effective density of states of the $\alpha$ movers (or the $\alpha$-injectivity
\cite{bufri2})
at the position $x_j$ of the wire, which contains the contribution of the local density of
states of the homogeneous wire, $\rho_{0,\alpha } ( 0,\omega)$ plus a correction due to
the backscattering by the impurity.
It is important to note that $I_j(x_j) \propto  \lambda_B $, while the current through
the wire is $I \propto \lambda^2_B $ (see also Refs. \cite{egger,dolc}).
For $T=0$ and low $V$, the condition $I_j=0$
leads to
\begin{equation} \label{mujinp}
\mu_j= \mu+\frac{V}{2}
\frac{ \rho^{eff}_{l}(x_j,\mu)-\rho^{eff}_{r}(x_j,\mu)}
{ \rho^{eff}_{l}(x_j,\mu)+ \rho^{eff}_{r}(x_j,\mu)},
\end{equation}
where we have made use of the fact that $\rho_j(\omega)$ is constant
and that $\rho^{eff}_{\alpha}(x_j,\omega) \sim
\rho^{eff}_{\alpha}(x_j,\mu)$ within a small window $|\omega
-\mu|\leq V/2$. Eq. (\ref{mujinp}) reduces to Eq.(37) of Ref.
\onlinecite{bufri2} for non-interacting electrons. It explicitly
shows that the local  potential monitors the difference between the
left and right injectivities  at the point $x_j$, relative to the
total
 density
of states at the given point. It is natural to expect that such an
observable should provide valuable information on the Friedel
oscillations introduced by the impurity \cite{egger,dolc,bufri1,bufri2},
as well as on the strength of the e-e interactions. In fact, a
low-energy expansion of the spectral densities casts:
\begin{eqnarray} \label{pl}
\mu_j & \sim & \mu+ C_1\,\lambda_B \, \sin(2 k_F x)\,  V^{2 \gamma +1}, \;\;\;T=0 \nonumber \\
\mu_j & \sim & \mu+ C_2 \,\lambda_B \,V \sin(2 k_F x)\,  T^{2
\gamma}, \;\;\;V \ll T,
\end{eqnarray}
with $C_1, C_2$ functions of $\gamma$.

The results of the full numerical calculation of  $\mu_j$ from
 the condition of $I_j(x_j)=0$ for arbitrary voltage differences and
temperature, with the exact effective density
 $\rho^{eff}_{\alpha}(x_j,\omega)$ (also evaluated numerically from eq.(\ref{rhoef})),
are shown in Fig. \ref{fig2}. At $T=0$, $\mu_j$ as a function of
the distance to the impurity, shown in the lower panel, oscillates
with the period $2k_F$ of the  Friedel oscillations, with a
voltage-dependent amplitude that follows the power law (\ref{pl})
(see upper left panel). For large distances to the impurity and
high $V$, beyond the scope of the approximations leading to
eqs. (\ref{pl}), the pattern shows additional structure, and the
amplitude of the oscillations decreases with the distance to the
impurity. The evolution of $\mu_j$ as the temperature increases,
corresponding to two values of the voltage, is illustrated in the
upper right panel of Fig. \ref{fig2}, where the dependence (\ref{pl})  is also
verified within the low $T$ and low $V$ regime, with $V \ll T$.
From these features we can infer the behavior of $R_{4pt}/R_{2pt}$ along
the sample by simply substituting (\ref{mujinp}) in (\ref{r4pt}). In
particular, we conclude that, as a function of $x$, $R_{4pt}/R_{2pt}$
should follow the pattern of Friedel oscillations, being
 positive or negative, depending on the points at which the probes are connected.
 As a function of $V$ it should be a power law with exponent $2 \gamma $.
As a function of $T$,
it should present rapid changes within the range $T<V$ and a crossover
to a power law with exponent $2 \gamma$ at higher temperatures.

We now consider the second situation of interest: the wire without
impurities ($\lambda_B=0$) but disordered probe leads. Thus,
the expressions for the Green's function of the wire
reduce to the ones for the homogeneous LL while the expression
for the tunneling current of eq.
(\ref{ijp}) reduces to eq. (\ref{ij01}) with $\rho_{\alpha}^{eff}(x_j,\omega) \equiv
\rho_{\alpha,0}(0,\omega)$.
While perfect
metallic systems are expected to have approximately flat densities
of states, impurities introduce effective
barriers, generating peaks in the
densities of states
 $\rho_j(\omega)$.

\begin{figure}
\includegraphics[width=0.9\columnwidth,clip]{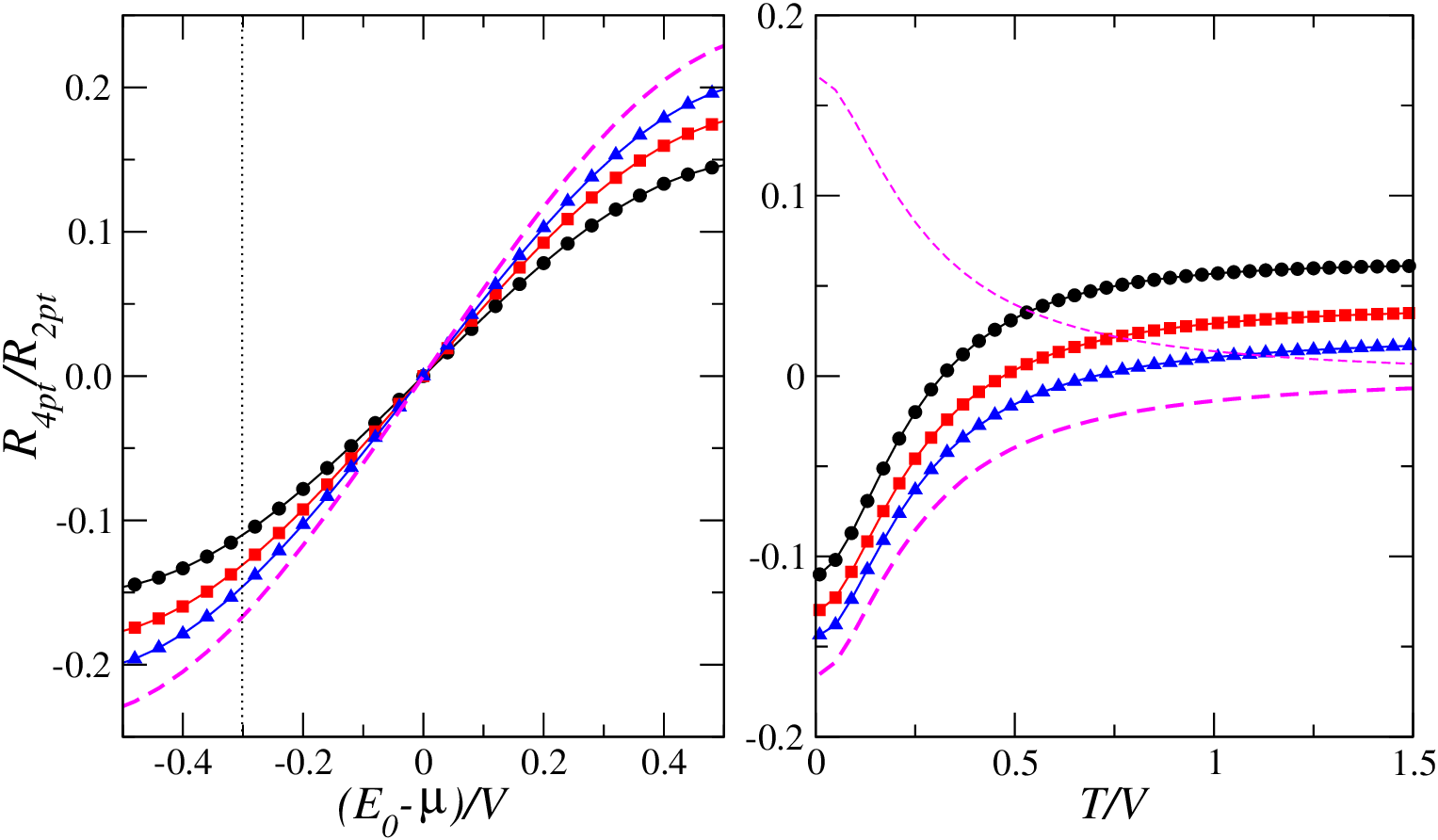}
\caption{\label{fig3} Four point resistance $R_{4pt}/R_{2pt}$ of a
LL with interaction parameter $K$ for a bias
$\mu_r-\mu_l=V$. The probe $j=1$ has a Lorentzian density of
states (a resonance) with width $\Delta=0.4V$ centered at $E_0$.
The probe $j=2$ has a constant density of states (which implies
$\mu_2=\mu=(\mu_l+\mu_r)/2$). The left panel corresponds to
temperature $T=0$ and shows the changes in $R_{4pt}$ as the center
of the resonance is moved around $\mu$. Circles, squares and
triangles correspond to $K=0.4,0.5,0.6$, while dashed lines, to
the non-interacting case ($K=1$). The right panel shows the
evolution with the temperature for the value of $(E_0-\mu)/V=-0.3$
indicated with dotted lines in the left panel. The thin dashed
line corresponds to $K=1$ and $(E_0-\mu)/V=0.3$.}
\end{figure}

Let us first analyze  $T=0$ and  probes with asymmetric density of
states, such that $\rho_j(\omega) \sim \rho^-_j(\mu), \mu-V/2 \leq
\omega \leq \mu$ and   $\rho_j(\omega) \sim \rho^+_j(\mu), \mu
\leq \omega \leq \mu+V/2 $. Then, the condition of a vanishing
current (\ref{ijp}) leads to:
\begin{eqnarray} \label{mjp}
\mu_j=\mu+\frac{V}{2} [\frac{  \rho^+_j(\mu)^{\nu} -  \rho^-_j(\mu)^{\nu}}
{\rho^-_j(\mu)^{\nu} +  \rho^+_j(\mu)^{\nu}}],
\end{eqnarray}
being $\nu=1/(2\gamma+1)$. For probes with symmetric
densities of states, we get $\mu_j=\mu$ and $R_{4pt}=0$. Instead, for an
asymmetric density of states, the potential drop between the
highest potential $r$ and the probe $j$ is lower (higher) than the
one between $j$ and $l$ for $\rho^-_j(\mu)>\rho^+_j(\mu)$
($\rho^-_j(\mu)<\rho^+_j(\mu)$), respectively, which reflects the
fact that the larger the spectral weight of the probe, the larger
the ability of that element to introduce resistive effects. For
finite temperature and very low voltage such that $T \gg V$, it
can be verified that $\mu_j = \mu, j=1,2$ and $R_{4pt}=0$.

Therefore, asymmetric densities of states of at least one of the
probes together with the condition $\rho_1(\omega) \neq
\rho_2(\omega)$ would lead to  a non vanishing $R_{4pt}$ when $T <
V$. An example is analyzed in Fig. \ref{fig3}. We
consider a Breit-Wigner model for
one of the probes, assuming
a single resonance within the window of width $V$
centered around $\mu$: $\rho_1(\omega) =A_1
/[(\omega-E_0)^2+\Delta]$, and a constant density of states
$\rho_2= A_2 $ for the other probe, where $A_1,A_2$ are
normalization constants and $\mu-V/2 \leq E_0 \leq  \mu+V/2$.
Under these conditions $\mu_2=\mu$, while $\mu_1$ is determined to
satisfy $I_1=0$.
 Results for
the corresponding relative resistance
$R_{4pt}/R_{2pt}=(\mu_1-\mu_2)/V$ are shown in Fig. \ref{fig3}.
The left panel corresponds to temperature $T=0$. When the center
of the resonance $E_0$ coincides with the mean chemical potential
$\mu$, the spectral weight spreads out symmetrically around this
point. Thus, $\rho_1^+(\mu)=\rho_1^-(\mu)$ and
 $\mu_1=\mu$, then $R_{4pt}=0$. As the center of the resonance moves to lower energies,
so does $\mu_1$ and $R_{4pt}$ becomes negative. Conversely, for
$E_0>\mu$, it is obtained $R_{4pt}>0$. Remarkably, interactions
tend to mask the structure observed in the non-interacting case
(with $K=1$). The behavior of $R_{4pt}$ as a function of the
temperature is shown in the right panel for the case
$(E_0-\mu)/V=-0.3$. Notice that the cases with $(E_0-\mu)/V > 0$
can be obtained from the ones with $(E_0-\mu)/V < 0$ by simply
transforming $R_{4pt} \rightarrow -R_{4pt}$ in the figure. In all the cases,
there is a range of temperature $T<V $, where $R_{4pt}$
experiments significant changes.

To conclude, let us comment on the theoretical and experimental
impact of our results. When an impurity is in the wire, it
 induces Friedel oscillations that manifest themselves in the local voltage and $R_{4pt}$.
The interactions leave a clear signature in the power law behavior
$R_{4pt}/R_{2pt} \propto V^{2 \gamma } $ and $R_{4pt}/R_{2pt}
\propto T^{2 \gamma}$. Interestingly, the exponent is different
from the one predicted in Ref. \onlinecite{dolc} for the two
terminal conductance of a LL with an impurity. This result has a
significant conceptual weight since it constitutes a concrete
example of the fact that different fundamental processes
contribute to each of these quantities ($R_{4pt}/R_{2pt} \propto
\lambda_B$ while $G \propto \lambda_B^2$). Therefore, a genuine
multiterminal setup is essential to evaluate $R_{4pt}$. For
impurities in the probes and an asymmetric configuration,
$R_{4pt}$ is determined by the way  in which the density of states
of the probe is distributed within an energy window of width $V$
centered in $\mu$,
 while the e-e
interactions play a milder role. In both cases, the behavior of
$R_{4pt}$ as a function of temperature at a sizable $V$, is
highly non universal and exhibits significant changes in the range
$T <V $. These results should help to provide a theoretical
framework to further analyze experimental data in  SWNT, like
those of Ref. \onlinecite{nano4pt} as well as to guide additional
 experiments along that line in the future.

We thank A.Bachtold, M.B\"uttiker and C.Chamon for useful discussions.
We acknowledge support from CONICET, Argentina,
 FIS2006-08533-C03-02 and  PIP-6157; PIA-11X386 UNLP, Argentina;
the ``RyC'' program from MCEyC, grant DGA for Groups of Excellence
 and AUIP of Spain.

\end{document}